\documentclass[12pt]{article}

\usepackage{authblk}
\usepackage{cite}
\usepackage{axodraw2}
\usepackage{amsmath,amsfonts}
\usepackage{amssymb}
\usepackage{macros}
\usepackage{caption}
\usepackage{subcaption}
\usepackage{bbm}
\usepackage{braket}
\usepackage{slashed} 
\usepackage{fullpage}
\usepackage{titlesec}

\titleformat{\section}
  {\normalfont\large\scshape}{\thesection}{1em}{}
\titleformat{\subsection}
  {\normalfont\scshape}{\thesubsection}{1em}{}

\renewcommand{\Im}{\operatorname{Im}}

\newcommand{\tB}{\tilde B}
\newcommand{\ms}{\mathbb} 

\newcommand{\U}{\ms U}

\renewcommand{\B}{\ms B}

\newcommand{\lf}{{\ell}}
\newcommand{\llf}{{\ell^2}}

\newcommand{\m}{\mu}
\newcommand{\n}{\nu}
\renewcommand{\t}{\tau}
\renewcommand{\r}{\rho}
\renewcommand{\l}{\lambda}
\renewcommand{\a}{\alpha}
\renewcommand{\b}{\beta}
\renewcommand{\d}{\delta}
\renewcommand{\g}{\gamma}
\renewcommand{\k}{\kappa}
\newcommand{\eps}{\epsilon}

\newcommand{\D}{\Delta}

\newcommand{\Kfree}{K_{\rm free}}

\title{\sc The Heat Kernel in Riemann Normal Coordinates and Multiloop Feynman Graphs in Curved Spacetime}
\author[1]{Igor Carneiro}
\author[1]{Gero von Gersdorff}
\affil[1]{Pontifícia Universidade Católica do Rio de Janeiro, Rio de Janeiro, Brazil}

\date{}
\begin{document}

\maketitle
\abstract{We present a formalism for computing arbitrary multi-loop Feynman graphs in curved spacetime using the heat kernel approach. To this end, we
 compute the off-diagonal components of the heat kernel in Riemann normal coordinates up to second order in the curvature.  }

\section{Introduction}

The heat kernel (HK) is an important tool in quantum field theory that has found numerous applications, including the calculation of the effective action, renormalization group equations of local operators, effective field theory of heavy particles, anomalies, chiral Lagrangians, to name a few. Moreover, it is  indispensable for  computing the behavior of a quantum field theory in classical gauge or gravitational backgrounds.
One of the  advantages of the method is that it is formulated in a manifestly gauge and diffeomorphism invariant way. 

Most of the applications of the method in the above contexts have focused on the one-loop case\cite{schwinger1951,dewitt1965dynamical,DeWitt:1967ub,Gilkey:1975iq,Barvinsky:1985an,Avramidi:1990ug,Avramidi:1990je,Fujikawa:1979ay,Fujikawa:1980eg,Gasser:1983yg,Ball:1988xg,vonGersdorff:2003dt,vonGersdorff:2006nt,Hoover:2005uf,Barvinsky:2005qi,vonGersdorff:2008df}
with few exceptions in the flat \cite{Duff:1975ue,Batalin:1976uv,Batalin:1978gt,Bornsen:2002hh} and curved \cite{Luscher:1982wf,Kodaira:1985pg,vandeVen:1991gw,Bilal:2013iva} backgrounds. 
However, a systematic formalism that enables one to write the local expansion of an arbitrary multi-loop Feynman graph is still missing. Recently such a formalism was developed for the flat case \cite{vonGersdorff:2022kwj,vonGersdorff:2023lle}, paving the way to writing the multi-loop effective action of an arbitrary quantum field theory in a fully gauge-covariant way. In this formalism one represents propagators by  HKs
\be
 K(t,x,x')=e^{-it(D^2+X)}\delta(x-x')
\ee
via $-i(D^2+X+m^2)^{-1}=\int_0^\infty dt\, K(t)e^{-it m^2}$, where $D$ is the gauge covariant derivative and $X$ is a background-field dependent mass. Notice that except in special cases it is necessary to know the off-diagonal ($x\neq x'$) part of the HK, as a generic diagram contains several  vertices connected by the propagators/HKs.
Knowledge of the off-diagonal HK is equivalent to the knowledge of the covariant derivatives of the HK at coincidence $x=x'$ \cite{Decanini:2005gt,Groh:2011dw}.\footnote{It is worth pointing out that even in the simple one-loop case, the diagonal HK is often insufficient, as a single loop can contain fields of different mass, spin etc.}

In this paper we generalize the formalism of refs.~\cite{vonGersdorff:2022kwj,vonGersdorff:2023lle} to curved space.
As we will see, a very convenient approach for this purpose is to adopt Riemann normal coordinates (RNCs) relative to some conveniently chosen base point, and expand the HK in {\em both} arguments $x$ and $x'$ in a conventional Taylor series.\footnote{RNCs have been applied previously in similar calculations, see e.g.~\cite{Luscher:1982wf,Larue:2023uyv}.} The resulting expansion coefficients are then given by 
(covariant derivatives of) curvature tensors at the base point. 
We will show that in RNCs the flat-space formalism generalizes straightforwardly to the curved background.

This paper is organized as follows. In section \ref{sec:general} we lay out our general formalism and derive our master formula, eq.~(\ref{eq:master3}). Section \ref{sec:hk} contains our main technical result, the fully off-diagonal HK up to dimension-four operators in RNCs. As an illustrative example for our formalism,  we compute in section \ref{sec:example} the 2-loop beta function of the $R\phi^2$ coupling of a scalar field in $\phi^4$ theory. Section  \ref{sec:conclusions} contains our conclusions. 
Our conventions are summarized in appendix \ref{sec:conventions}, and some well-known RNC expansions of geometric quantities are reproduced in appendix \ref{sec:rnc}. Finally, in appendix \ref{sec:jacobian} we give an alternative formulation that treats all vertices of a given graph in a symmetric way.

\section{General Formalism}
\label{sec:general}

In refs~\cite{vonGersdorff:2022kwj,vonGersdorff:2023lle} the following  formula for the contribution of a given Feynman graph to the effective Lagrangian in flat space was derived:\footnote{Throughout this paper, we use the shorthands $\int_{\t_i}=\prod_{i=1}^P\int_0^\infty d\t_i$ and $\int_{x_n}=\prod_{n=1}^V\int_{} d^dx_n$.}
\be
\mathcal L_{\rm eff}(x)= \int_{\t_i}
\bigl [ I(\t_i,\,i\partial_{n},\,-i\partial_{i})
\,\Gamma(\t_i,\,x+y_n,\,k_i)
\bigr]_{y_n=0,\ k_i=0}\,.
\label{eq:master}
\ee
Here $\t_i$ are Schwinger parameters (one for each edge of the graph), and $I(\t_i,\,p_n,\,z_i)$ is the Fourier transform of 
\be
I(\t_i,\,y_n,\,k_i)\equiv 
\delta(\bar y_0)\,i^{-L}
\exp(\t_j k_j^2+iy_m \B_{mj} k_j)\\
\equiv 
\delta(\bar y_0)\hat I(\t_i,\,y_n,\,k_i)
\label{eq:IhatI}
\ee
with respect to both vertex positions $y_n$ and edge momenta $k_i$. Here
 $\B$ is the (directed) incidence matrix of the graph, and
\be
\bar y_0\equiv \frac{1}{V}\sum_{n=1}^Vy_n\,,
\label{eq:ybar}
\ee
 is the "center of mass" of the vertices.
The function $I(\t_i,\,p_n,\,z_i)$ can be given in terms of four graph polynomials
and has been studied in great detail in ref \cite{vonGersdorff:2023lle}. 
The function $\Gamma(\t_i,\,x_n,\,k_i)$  contains the dependence on the background fields. 
Besides  the vertex positions $x_n=x+y_n$ it also depends on the edge momenta $k_i$ which naturally appear in the presence of fermion propagators or propagators of derivative of fields. Here, we will exclusively be studying scalar fields without derivative interactions. In this case, $\Gamma$ does not depend on the $k_i$ and we  have simply 
\be
\Gamma(\t_i,x_n)\equiv \prod_n C_n\prod_j B_j\, e^{-\t_j m_j^2}
\label{eq:defGamma}
\ee
where the $m_i$ are masses of the $i$th propagator, $C_n$ are field-dependent couplings (one for each vertex $x_n$) and $B_j$ denotes the rescaled HK of the $j$th propagator,
\be
B(\t,x,x')\equiv \frac{K(-i\t,x,x')}{\Kfree(-i\t,x,x')}
\ee
with 
\be
\Kfree(t,x,x')\equiv e^{-it\partial^2}\delta(x-x')
=
i(4\pi it)^{-\frac{d}{2}}
e^{-i\frac{(x-x')^2}{4t}}\,.
\label{eq:kfree}
\ee
Since in the pure scalar case $\Gamma$ is independent of the momenta $k_i$, the function $I(\t_i,\,p_n,\,z_i)$ is  only needed at the arguments $z_i=0$, in which case it is given by
\be
I(\t_i,\,p_n)\equiv I(\t_i,\,p_n,\,0)=
(4\pi)^{-\frac{{dL}}{2}}
\D^{-\frac{d}{2}}
\exp\left(- \U_{mn} p_mp_n
\right)\,,
\label{eq:defI2}
\ee
where $\Delta(\t_i)$ is the first Szymanzik polynomial of the graph and $\U(\t_i)$ is a $V\times V$ square matrix in terms of which the  second Szymanzik polynomial reads $\Delta( \U_{mn}p_mp_n +m_i^2\t_i)$.
The integration over $k_i$ at $z_i=0$ is essentially the usual momentum integration of the graph.
The pure scalar case with no derivative interactions can therefore  be written as
\be
\mathcal L_{\rm eff}(x)= \int_{\t_i}
\bigl [ I(\t_i,\,i\partial_{n})
\,\Gamma(\t_i,\,x+y_n)
\bigr]_{y_n=0}\,,
\label{eq:master2}
\ee
with $\Gamma$ and $I$  given in eqns.~(\ref{eq:defGamma}) and (\ref{eq:defI2}).
In evaluating eq.~(\ref{eq:master2}) one needs to know the off-diagonal part of the function $B$, or, equivalently, its various covariant derivatives at coinciding arguments.

We now want to generalize this formalism to curved spacetime.
To keep it simple, we consider only scalars and also  ignore any gauge background.
Both nonzero spin and gauge dependent backgrounds can straightforwardly be dealt with, we will postpone this to future work.
We start out by writing the contribution to the effective action resulting from a specific Feynman graph in an obvious way as 
\be
S_{\rm eff}=i^{-L}\int_{\t_i}\int_{x_n}F(\t_i,\,x_n)
\label{eq:Seff0}
\ee
where
\be
F(\t_i,\, x_n)\equiv \prod_n C_n\prod_j K_j\, e^{-\t_j m_j^2}
\ee
The field-dependent coupling of $N$ (fluctuation) fields is normalized as
\be 
C\equiv \sqrt{g}\frac{\delta^N}{\delta \phi^N}\mathcal L_{\rm int}
\ee
where $\mathcal L_{\rm int}$ is the Lagrangian containing the interaction.
The HKs $K_j$ transform as a scalar at both arguments. For simplicity, in eq.~(\ref{eq:Seff0}) we have suppressed symmetry factors and signs from closed fermion loops.
Next we use $1=\int_x\sqrt{g(x)}\, \delta(x,\tilde x)$ to write 
\be
S_{\rm eff}=i^{-L}\int_{\t_i}\int_x \sqrt{g(x)}\int_{x_n} \delta(x,\tilde x)F(\t_i,\,x_n)\,.
\label{eq:Seff}
\ee
where $\delta(x,\tilde x)=(g(x)g(\tilde x))^{-\frac{1}{4}}\delta(x-\tilde x)$ is  the scalar delta function.
Eq.~(\ref{eq:Seff}) is  valid for an arbitary function $\tilde x(x_n)$ of the vertex positions, we will make a convenient choice below.
The (scalar) effective Lagrangian thus reads
\be
\mathcal L_{\rm eff}(x)=i^{-L}\int_{\t_i}\int_{x_n} \delta(x,\tilde x)F(\t_i;x_n)
\ee
We now make a change of integration variables  $x_n\to y_n$ where the $y_n$ are RNCs with base point $x$ which we conveniently chose to coincide with the origin $y=0$.
We review RNCs in appendix \ref{sec:rnc}.
The most important aspect of RNCs is that the metric and hence all geometric objects such as connection, volume form, HK etc, have a conventional Taylor expansion about the origin, with coefficients given by curvature tensors and their covariant derivatives.

In particular, we compute in section \ref{sec:hk} the expansion of the HK up to dimension four operators in the background fields. Up to dimension two it reads:
\be
B(\t,y,y')=
1-\tfrac{\t}{6}R-\t X+\tfrac{1}{12}R_{\mu\nu}(y-y')^\mu(y-y')^\nu-\tfrac{1}{12\t}R_{\mu\nu\rho\sigma}y^\mu y'^\nu y^\rho y'^\sigma+\dots
\label{eq:hkoff}
\ee
where, as before, $B=K/\Kfree$ and the fields on the right hand side are evaluated at $y=0$. This result contains the usual diagonal HK coefficients for $y=y'$, as well as the off-diagonal ones for $y'=0$ \cite{Luscher:1982wf}  as special cases. 
For Feynman graphs with  three or more vertices one  needs the most general off-diagonal HK of eq.~(\ref{eq:hkoff}).
Notice that as with any RNC expansion, the coefficients are tensors at the base point $x$.
The unusual appearance of negative powers of $\t$ is due to the fact that we normalized the HK by $K_{\rm free}\sim e^{-i(y-y')/4t}$ instead of the usual $e^{-i\sigma(y,y')/2t}$ factors. The additional expansion of Synge's world function $\sigma(y,y')$ in RNCs then generates negative powers of $\t$. However, these terms only enter in the effective action via application of the derivative operator eq.~(\ref{eq:master3}), and these derivatives are companied by additional  positive powers of the $\t_i$.
 
We next turn to the quantity $\tilde x$ in eq.~(\ref{eq:Seff}), which  can be any function of the $V$ vertex positions $x_n$. In the flat case, we chose the center of mass position of the vertices. 
While it is possible to generalize the definition eq.~(\ref{eq:ybar}) to curved space (see appendix \ref{sec:jacobian}), the  less symmetric choice $\tilde x=x_1$ (or similarly any other vertex) is simpler. In RNCs with base point $x$, $\delta (x,x_1)$  becomes $\delta(y_1)$. 
Putting everything together,
the factors $i^{-L}$, $\prod_j K_{\rm free, j}$ and $\delta(y_1)$ combine to give a function 
$I'(\t_i,y_n,k_i)$, obtained from eq.~(\ref{eq:IhatI}) by replacing $\delta(\bar y_0)$ by $\delta(y_1)$. This leads to a function $I'(\t_i,p_{n'})$, that is  is related to the function $I(\t_i,p_n)$ of eq.~(\ref{eq:defI2}) by momentum conservation:\footnote{One can also give $I$ in terms of $I'$ as $I(\t_i,p_n)=I'(\t_i,p_{n'}-\bar p)$ where 
$\bar p=\frac{1}{V}\sum_{n=1}^V{p_n}$.}
\be
I'(\t_i,p_{n'})
=I(\t_i,p_{n})|_{p_1=-(p_2+\dots+p_V)}
\label{eq:Iprime}
\ee
The primed indices $n'$ only run over $2\dots V$, in particular, $I'(\t_n, p_n)$ is independent of $p_1$.
It is then clear that our master formula eq.~(\ref{eq:master2}) continues to hold in RNCs, 
\be
\mathcal L_{\rm eff}=\int_{\t_i} I'(\t_i,i\partial_{n'})\Gamma(\t_i,y_{n})|_{y_{n}=0}
\label{eq:master3}
\ee
%
where the expansion of $\Gamma$ in RNCs can be obtained from the expansion of its factors ($C_n$ and $B_j$).

\section{The heat kernel in Riemann normal coordinates}

\label{sec:hk}

In this section we calculate the off-diagonal HK in RNCs, keeping 
terms up to dimension four. 

We choose a base point on the manifold, $x$, and RNCs $y^\mu$ with $x$ having coordinates $y^\mu=0$.
Consider the scalar Lorentzian HK 
\be
K(t,y,y')=e^{-it (\nabla^ 2+X)}\delta(y,y')
\ee
where $\delta(y,y')=[g(y)g(y')]^{-\frac{1}{4}}\delta(y-y')$ is the scalar delta function and $g\equiv -\det g$.
Let us temporarily parametrize the HK as
\be
K(t,y,y')\equiv \tilde B(it,y,y')i(4\pi it)^{-\frac{d}{2}}e^{-i\frac{\sigma(y,y')}{2t}}
\ee
The relation to $B$ as defined in section \ref{sec:general} is 
\be
B(\t,y,y')= e^{\frac{1}{2\t}\left(\sigma(y,y')-\tfrac{1}{2}(y-y')^2\right)}\tilde B(\t,y,y')
\label{eq:BBt}
\ee
Unlike  $B$, the function $\tilde B$ transforms as a biscalar. The exponential prefactor equals one in the flat limit as well as the limits $y\to 0$ or $y'\to 0$.
Let us define the following notation.
\bea
[\tilde B_{;\mu\dots;\nu\dots}]_{;\rho\dots}\equiv \lim_{y\to 0} 
(\cdots\nabla_\rho)\lim_{y'\to y}(\cdots \nabla_\mu)(\cdots\nabla'_\nu) \tilde B(y,y')
\eea
A similar notation holds for partial derivatives, which are denoted by a comma instead of a semicolon. 
Notice in particular the final limit of $y\to 0$.
In terms of this notation, the expansion coefficients we are looking for are given by $[\tilde B_{,\mu\nu\dots,\rho\sigma\dots}]$ and 
$[B_{,\mu\nu\dots,\rho\sigma\dots}]$.
The coefficients with more than four derivatives on $\tilde B$ only contribute operators fo dimension 5 or higher, so we can disregard them.
Due to the symmetry 
$\tilde B(y,y')=\tilde B(y',y)$ we have relations such as $[\tB_{,\mu\nu,\r}]=[\tB_{,\r,\mu\nu}]$ etc, so we need to only find the coefficients 
$[\tilde B]$, $[\tB_{,\m}]$, $[\tB_{,\m\n}]$, $[\tB_{,\m\n\r}]$, $[\tB_{,\m\n\r\s}]$, 
$[\tB_{,\m,\n}]$,
$[\tB_{\m\n,\r}]$,
$[\tB_{,\m\n\r,\s}]$, and $[\tB_{,\m\n,\r\s}]$.

%

We can related the expansion coefficients to {\em symmetrized covariant derivatives} via
\begin{align}
[\tB_{,\m}]&=[\tB_{;\m}]\nn\\
[\tB_{,\m\n}]&=[\tB_{;\m\n}]\nn\\
[\tB_{,\m\n\r}]&=[\tB_{;(\m\n\r)}]\nn\\
[\tB_{,\m\n\r\s}]&=[\tB_{;(\m\n\r\s)}]\nn\\
[\tB_{,\m,\n}]&=
	[\tB_{;\m}]_{;\n}
	-[\tB_{;\m\n}]
\nn\\
[\tB_{\m\n,\r}]&=
	[\tB_{;\m\n}]_{;\r}
	-[\tB_{;(\m\n\r)}]
	+{
	\Gamma^\a_{\m\n,\r}[\tB^{}_{;\a}]
	}
\nn\\
[\tB_{,\m\n\r,\s}]&=[\tB_{;(\m\n\r)}]_{;\s}
	-[\tB_{;(\m\n\r\s)}]
	+3[\tB^{}_{;\a(\m}]\Gamma^\a_{\n\r),\s}
	+ {\Gamma^\a_{(\m\n,\r)\s}[\tB^{}_{;\a}]}
\nn\\
[\tB_{,\m\n,\r\s}]&=
	[\tB_{;\m\n}]_{;\r\s}
	-[\tB_{;(\m\n\s)}]_{;\r}
	-[\tB_{;(\m\n\r)}]_{;\s}	
	+[\tB_{;(\m\n\r\s)}]
	\nn\\&
	-\Gamma^\a_{\m\s,\r}[\tB^{}_{;\a\n}]
	-\Gamma^\a_{\n\s,\r}[\tB^{}_{;\a\m}]
	-\Gamma^\a_{\m\n,\s}([\tB_{;\a \r}]-{[\tB_{;\a}]_{;\r}})
	-\Gamma^\a_{\m\n,\r}([\tB_{;\a \s}]-{[\tB_{;\a}]_{;\s}})
	\nn\\&
	+{
	 \tfrac{1}{3}\Gamma^\a_{\m\n,\r\s}[\tB^{}_{;\a}]
	-\tfrac{1}{3}\Gamma^\a_{\m\r,\n\s}[\tB^{}_{;\a}]
	-\tfrac{1}{3}\Gamma^\a_{\n\r,\m\s}[\tB^{}_{;\a}]
	-\tfrac{1}{3}\Gamma^\a_{\m\s,\n\r}[\tB^{}_{;\a}]
	-\tfrac{1}{3}\Gamma^\a_{\n\s,\m\r}[\tB^{}_{;\a}]
	}
\label{eq:expansioncoeff}	
\end{align}
where the parentheses on the indices denote symmetrization with strength one. 
These relations are easily verified by straightforward devovariantization.
Notice that the connection terms are evaluated at the origin, and hence can be read off from eq.~(\ref{eq:ChristoffelRNC}).\footnote{In particular, $\Gamma^\mu_{\r\s}(0)=0$.}
The quantities on the left hand side are the expansion coefficients  
of $\tilde B(\t,y,y')$ that we are looking for. 

The symmetrized covariant derivatives of $\tilde B$ at coincidence $y'=y$ can be computed by standard techniques, one finds  \cite{Decanini:2005gt,Groh:2011dw}
\begin{align}
\tilde B(\t,y,y)=&
	\ 1
	-\tfrac{\t}{6}R
	-\t X
	+\tfrac{\t^2}{30}\nabla^2 R
	+\tfrac{\t^2}{72}R^2
\nn\\&
	-\tfrac{\t^2}{180}R_{ij}^2
	+\tfrac{\t^2}{180}R_{ijkl}^2
	+\tfrac{\t^2}{6}\nabla^2 X
	+\tfrac{\t^2}{6}RX
	+\tfrac{\t^2}{2}X^2+\dots
\nn\\
\tilde B_{;\mu}(\t,y,y)=&
	-\tfrac{\t}{12}R_{;\mu}
	-\tfrac{\t}{2}X_{;\mu}+\dots
\nn\\
\tilde B_{;(\mu\nu)}(\t,y,y)=&
	\ \tfrac{1}{6}R_{\mu\nu}
	-\tfrac{\t}{20}R_{;\mu\nu}
	-\tfrac{\t}{60}\nabla^2R_{\mu\nu}
	-\tfrac{\t}{3}X_{;\mu\nu}
\nn\\&
	+\tfrac{\t}{45}R^\a_{\ \mu}R^{}_{\a\nu}
	-\tfrac{\t}{36}RR_{\mu\nu}
	-\tfrac{\t}{90}R^{\a\b}R_{\a\mu\b\nu}
	-\tfrac{\t}{90}R^{\a\b\g}_{\ \ \ \ \mu}R^{}_{\a\b\g\nu}
	-\tfrac{\t}{6}XR_{\mu\nu}+\dots
\nn\\
\tilde B_{;(\mu\nu\rho)}(\t,y,y)=&
	\ \tfrac{1}{4}R_{(\mu\nu;\rho)}+\dots
\nn\\
\tilde B_{;(\mu\nu\rho\sigma)}(\t,y,y)=&
	\ \tfrac{3}{10}R_{(\mu\nu;\rho\sigma)}
	+\tfrac{1}{15}R_{\a(\mu\nu}^{\ \ \ \ \ \b}R^\a_{\ \rho\sigma)\b}
	+\tfrac{1}{12}R_{(\mu\nu}R_{\rho\sigma)}+\dots
\label{eq:covariantsym}
\end{align}
where the ellipsis denotes operators of dimension five or higher.

Let us define $\tB(\t,y,y')|_D$ the part of $\tB$ that contains all dimension $D$ operators,
in particular, $\tB(\t,y,y')|_0=1$, $\tB(\t,y,y')|_1=0$. 
Using eq.~(\ref{eq:covariantsym}) in eq.~(\ref{eq:expansioncoeff}) yields
\begin{align}
\tilde B(\t,y,y')|_2=&
	- \tfrac{\t}{6}R-\t X
	+ \tfrac{1}{12}R_{\mu\nu} 
		\left(y^\m y^\n+y'^\m y'^\n\right)
	-\tfrac{1}{6}R_{\mu\nu} \,
		y^\m y'^\n
\\
\tilde B(\t,y,y')|_3=&	
	 \left(
	-\tfrac{\t}{12}R_{;\mu}
	-\tfrac{\t}{2}X_{;\mu}\right)
		\left(y^\m+y'^\n\right)
	+\tfrac{1}{24}R_{\mu\nu;\rho}\,
		\left(y^\m y^\n y^\r +y'^\m y'^\n y'^\r\right)
	\nn\\
	&+\left(\tfrac{1}{12}R_{\mu\nu;\rho}-\tfrac{1}{6}R_{\rho\mu;\n}\right)
		\tfrac{1}{2}\left(y^\m y^\n y'^\r+y'^\m y'^\n y^\r\right)
\\
%
\tilde B(\t,y,y')|_4 =&	
	\ \tfrac{\t^2}{30}\nabla^2 R
	+\tfrac{\t^2}{72}R^2
	-\tfrac{\t^2}{180}R_{\m\n}R^{\m\n}
	+\tfrac{\t^2}{180}R_{\m\n\r\s}R^{\m\n\r\s}
	+\tfrac{\t^2}{6}\nabla^2 X
	+\tfrac{\t^2}{6}RX
	+\tfrac{\t^2}{2}X^2
	\nn\\&
	+\left(	
	-\tfrac{\t}{20}R_{;\mu\nu}
	-\tfrac{\t}{60}\nabla^2R_{\mu\nu}
	-\tfrac{\t}{3}X_{;\mu\nu}
	-\tfrac{\t}{6}XR_{\mu\nu}
	+\tfrac{\t}{45}R^\a_{\ \mu}R^{}_{\a\nu}
	\right.
\nn\\&
	\phantom{+}
	\left.
	-\tfrac{\t}{36}RR_{\mu\nu}
	-\tfrac{\t}{90}R^{\a\g}R_{\a\mu\g\nu}
	-\tfrac{\t}{90}R^{\a\g\kappa}_{\ \ \ \ \mu}	R^{}_{\a\g\kappa\nu}
	\right)
		\tfrac{1}{2}\left(y^\mu y^\nu+y'^\m y'^\n\right)
	\nn\\
	&+\left(
	-\tfrac{\t}{30}R_{;\mu\nu} 
	+\tfrac{\t}{60}\nabla^2R_{\mu\nu}
	-\tfrac{\t}{6}X_{;\mu\nu}
	+\tfrac{\t}{6}XR_{\mu\nu}
	-\tfrac{\t}{45}R^\a_{\ \mu}R^{}_{\a\nu}
	\right.
	\nn\\&
	\phantom+
	\left.
	+\tfrac{\t}{36}RR_{\mu\nu}
	+\tfrac{\t}{90}R^{\a\g}R_{\a\mu \g\nu}
	+\tfrac{\t}{90}R^{\a\g\kappa}_{\ \ \ \ \mu}R^{}_{\a\g\kappa\nu}
	\right)
		y^\mu y'^\nu
	\nn\\	
	&+
	\left(	
	\tfrac{3}{10}R_{\mu\nu;\rho\sigma}
	+\tfrac{1}{15}R_{\a\mu\nu}^{\ \ \ \ \g}R^\a_{\ \rho\sigma\g}
	+\tfrac{1}{12}R_{\mu\nu}R_{\rho\sigma}\right)
		\tfrac{1}{24}\left(y^\m y^\n y^\r y^\s+y'^\m y'^\n y'^\r y'^\s\right)
	\nn\\
	&+\left(
	\tfrac{1}{120}R_{\mu\nu;\rho\sigma}
	-\tfrac{3}{20}R_{\s\m;\n\r}
	+\tfrac{11}{120}R_{\m\n;\s\r}
	\right.
	\nn\\
	&
	\phantom{+}
	\left.
	-\tfrac{1}{15}R_{\a\mu\nu}^{\ \ \ \  \g}R^\a_{\ \r\s \g}
	-\tfrac{1}{12}R_{\mu\nu}R_{\rho\sigma}
	\right)
		\tfrac{1}{6}\left(y^\m y^\n y^\r y'^\s+y'^\m y'^\n y'^\r y^\s\right)
	\nn\\
	&+
	\left(
	\tfrac{19}{180}R_{\m\n;\r\s}
	+\tfrac{19}{180}R_{\r\s;\m\n}	
	-\tfrac{11}{90}R_{\m \r;\n \s}
	-\tfrac{11}{90}R_{\r\m;\s \n}
 	- \tfrac{1}{18} \nabla^2 R_{\m\r\n\s}
	\right.\nn\\
	&\left.
	\phantom{+}
	-\tfrac{11}{90}R_{\m\r\a\g}R_{\n\s}^{\  \ \,\a\g}
	+\tfrac{7}{45}
	R_{\a\m\r}^{\ \ \ \  \g}R^\a_{\ \n\s \g}
	-\tfrac{4}{45}
	R_{\a\mu\nu}^{\ \ \ \  \g}R^\a_{\ \r\s \g}
	\right.\nn\\
	&\left.
	\phantom{+}
	+\tfrac{1}{36}R_{\m\n}R_{\r\s}
	+\tfrac{1}{18}R_{\m\r}R_{\n\s}
	\right)
		\tfrac{1}{4}y^\m y^\n y'^\r y'^\s
\label{eq:B4}
\end{align}
In the last two parenthesis in eq.~(\ref{eq:B4}), we have eliminated the terms with one Ricci and one Riemann tensor by use of Bianchi identities and commutators of covariant derivatives.

Finally, we switch from $\tilde B$ to $B$ using eqns.~(\ref{eq:BBt}) and (\ref{eq:SyngeRNC})
\begin{align}
B(\t,y,y')|_2=& 
	\ \tB(\t,y,y')|_2
	-\tfrac{1}{12\t}R_{\m\r\n\s}\, 
		y^\m y^\n y'^\r y'^\s 
\\
B(\t,y,y')|_3=& 
	\ \tB(\t,y,y')|_3
	-\tfrac{1}{24\t}R_{\m\r \n\s;\t}\,
		y^\mu y^\nu(y^\t+y'^\t) y'^\r y'^\s 
\\
B(\t,y,y')|_4=&
	\ \tB(\t,y,y')|_4
 	-
	\left(
	\tfrac{1}{3\t}R_{\m\r\n\s}R_{\k\t}
	\right.	
	\nn\\&	
	\left.
	+\tfrac{4}{15\t}R^{\alpha}_{\ \k\m\r}R_{\alpha\t\n\s}^{}
	+\tfrac{1}{5\t}R_{\mu\r\nu\s;\k\t}\right) 
		\tfrac{1}{48}\, y^\m y^\n (y^\k y^\t+y'^\k y'^\t)y'^\r y'^\s
	\nn\\&	
	+\left(
	\tfrac{1}{2\t}R_{\m\n\r\s}R_{\k\t}
	-\tfrac{4}{5\t}R^{\alpha}_{\ \mu\nu\t}R_{\alpha\r\s\k}^{}
	\right.	
	\nn\\&	
	\left.
	-\tfrac{3}{40\t}R_{\mu\r\nu\s;\k\t}
	-\tfrac{3}{40\t}R_{\mu\r\nu\s;\t\k}
	\right)
		\tfrac{1}{36}\,y^\mu y^\nu  y^\k y'^\r y'^\s y'^\t 
	\nn\\&
	+\tfrac{1}{144\t^2}R_{\m\r\n\s}R_{\t\k\l\eta}\,
		y^\m y^\n y^\t y^\l y'^\r y'^\s y'^\k y'^\eta
\end{align}

\section{Example}

\label{sec:example}

As a simple enough example to illustrate the formalism let us compute the two-loop beta function of the nonminimal scalar coupling to gravity  
\be
\mathcal L_{\xi}=-\frac{1}{2}\xi_0  R\phi_0^2
\ee
in $\phi^4$ theory. Here and in the following, the subindex $0$ denotes bare (unrenormalized) quantities. Since the fluctuations of the graviton are Planck suppressed we will ignore them  and focus on the scalar ones. Then we have the following background-field dependent mass
\be
X=\frac{\lambda_0}{2}\phi_0^2+\xi_0 R
\ee
as well as the triliner and quadrilinear field dependent couplings 
$C^{(N)}=\frac{\delta^N}{\delta\phi^N}\sqrt g\mathcal L$
\be
 C^{(3)}=-\sqrt{g}\lambda_0 \phi_0
\qquad C^{(4)}=-\sqrt{g}\lambda_0
\ee

Let us first consider the one-loop contribution to the effective action, which is not directly given by eq.~(\ref{eq:master3}), but by the standard formula in terms of the diagonal HK, which in our notation reads
\be
\mathcal L^{\rm 1-loop}_{\rm eff}=\frac{1}{2}\frac{1}{(4\pi)^{\frac{d}{2}}}\int_\t\frac{e^{-m_0^2 \t}}{\t^{\frac{d}{2}+1}}B(\t,0,0)
\ee
We can directly determine the effective Lagrangian given by this contribution. Considering only terms  proportional to one of the interactions in the original Lagrangian alongside the kinetic and mass contributions, one obtains, using the explicit expressions for $B$ form section  \ref{sec:hk},
\begin{align}
\mathcal L_{\rm eff}^{\rm 1-loop}=
\frac{1}{2}\frac{1}{(4\pi)^{\frac{d}{2}}}
\biggl\{\biggr.
&m_0^{d-2}\,\Gamma(1-\tfrac{d}{2})\left[-\tfrac{\lambda_0}{2}\phi_0^2\right]
\nn\\
+\,&m_0^{d-4}\,\Gamma(2-\tfrac{d}{2})\left[\tfrac{\lambda_0^2}{8}\phi_0^4+\tfrac{\lambda_0}{2}\left(\tfrac{1}{6}+\xi_0\right)R\phi_0^2\right]
\biggl.\biggr\}
\label{eq:1loop}
\end{align}

\begin{figure}
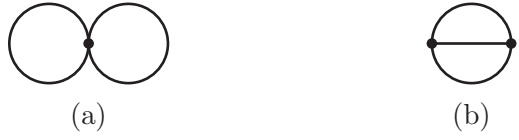

\begin{center}

\begin{subfigure}[b]{0.3\linewidth}
  \centering
  \begin{axopicture}(60,30)(0,0)
    \SetWidth{1.0}
   	\Vertex(30,15){2}
	\Arc(15,15)(15,0,360)
	\Arc(45,15)(15,0,360)
\end{axopicture}
\label{fig:fig8}
\caption{}
\end{subfigure}
\begin{subfigure}[b]{0.3\textwidth}
	\centering
	\begin{axopicture}(30,30)(0,0)	
	\SetWidth{1.0}
    \SetColor{Black}
	\Arc(15,15)(15,0,360)
	\Line(0,15)(30,15)
	\Vertex(0,15){2}
	\Vertex(30,15){2}
\end{axopicture}
\caption{}
\label{fig:sunset}
\end{subfigure}
\caption{Two-loop diagrams.}
\label{fig:diagrams}
\end{center}
\end{figure}

The two-loop contribution to the effective action is given by the diagrams in figure \ref{fig:diagrams}.
We now apply our master formula eq.~(\ref{eq:master3}), which we recall is derived by setting one of the interaction points as the RNC origin. This choice greatly simplifies the calculation of the sunset diagram by having to consider only the derivative of $\Gamma$ with respect to one of the two vertices.  
The contributions arising from the figure-8 and sunset diagrams can be directly computed from eq.~(\ref{eq:master3}) by using the appropriate \(\Delta\) and \(\mathbb{U}\) definitions for the respective graph in eq.~(\ref{eq:defI2}) and eq.~(\ref{eq:Iprime}), as well as including their symmetry factors (respectively, \(\frac{1}{8}\) and \(\frac{1}{12}\)). For the figure-8, the Szymanzik polynomial is simply given by the product of the  Schwinger parameters \(\tau\) and \(\sigma\) and the matrix \(\mathbb{U}\) vanishes. Hence,
\begin{align}
\label{eq:fig8}
\mathcal L_{\rm eff}^{\rm fig-8}
&=
\frac{1}{8}\frac{1}{(4\pi)^d}\int_{\t,\s}
\frac{e^{-m_0^2(\t+\s)}}{(\t\s)^{\frac{d}{2}}}\left\{-\lambda_0 B(\t,0,0)B(\s,0,0)\right\}
\\
&
\begin{aligned}
{}=-\frac{\l_0}{8}\frac{1}{(4\pi)^d}\biggl\{\biggr.
&m_0^{2d-6}\Gamma(1-\tfrac{d}{2})
\Gamma(2-\tfrac{d}{2})\left[-\tfrac{\l_0}{2}\phi_0^2\right]
\nn\\
+\,&m_0^{2d-8}
\left(\Gamma(2-\tfrac{d}{2})^2+\Gamma(1-\tfrac{d}{2})\Gamma(3-\tfrac{d}{2})\right)\left[\tfrac{\l_0^2}{4}\phi_0^4+\l_0\left(\tfrac{1}{6}+\xi_0\right)R\phi_0^2\right]
\biggl.\biggr\}
\end{aligned}
\end{align}
As expected from the topology of the diagram, only the diagonal HK contributes.

Finally, the sunset diagram's first Szymanzik polynomial is given by the  sum of all possible products of two Schwinger parameters, $\D=\t_1\t_2+\t_1\t_3+\t_2\t_3$. It also has a non-vanishing $\U$ matrix  given by eq.~(5.32) of \cite{vonGersdorff:2023lle}, leading to 
$I'(\t_i,p_{n'})=(4\pi)^{-d}\Delta^{-\frac{d}{2}}\exp(\frac{\t_1\t_2\t_3}{\D}p_2^2)$. Hence
\begin{align}
\mathcal L_{\rm eff}^{\rm sunset}&=\frac{1}{12}\frac{1}{(4\pi)^d}
\int_{\t_i}\frac{e^{-m_0^2\sum_i\t_i}}{\D^\frac{d}{2}}
\left[
e^{-\frac{\t_1\t_2\t_3}{\D}\partial_y^2}\, \l_0^2\,\phi_0(0)\phi_0(y)\sqrt{g(y)}
\prod_{i=1}^3 B(\t_i,0,y)\right]_{y=0}
\nn\\
&=\frac{\l_0^2}{12}\frac{1}{(4\pi)^d}\biggl\{ J^\frac{d}{2}_{111}\,\phi_0^2
	+J_{222}^{\frac{d}{2}+1}\left[-\tfrac{1}{6}R-\phi_0\nabla^2\phi_0\right]+3J_{211}^\frac{d}{2}\left[-\left(\tfrac{1}{6}+\xi_0\right)R\phi_0^2-\tfrac{\lambda_0}{2}\phi_0^4\right]\biggr\}
	\label{eq:sunset}
\end{align}
where $J_{\nu_1\nu_2\nu_3}^\nu\equiv \int_{\t_i}\Delta^{-\nu}\prod_i(\t_i^{\nu_i-1}e^{-m_0^2\t_i})$ are the (proper) two-loop vacuum integrals in Schwinger parametrization.
In the last line of eq.~(\ref{eq:sunset}), the middle term in the curly bracket stems from the $\mathcal O(y^2)$ terms in the RNC expansions of $B(\t,0,y)$ (see section \ref{sec:hk}), as well as $g(y)$ and $\phi_0(y)$ (see appendix \ref{sec:rnc}). Notice that we already used $g(0)=1$ at the first vertex.
While the one-loop and figure-8 contributions only required the knowledge of the 
diagonal HK, the sunset contribution also require the (partially) off-diagonal one $B(\t,0,y)$. At three loop order, the full off-diagonal HK $B(\t,y,y')$ would be needed (see below), and the same is true for one and two loop contributions in more complicated theories (for instance, once we include fluctuations of the graviton field).

We now write \(d=4-\epsilon\), and take the limit \(\epsilon\rightarrow 0\), keeping only singular terms necessary in the calculation of the counterterms and beta functions. The singularities of the 1-loop and figure-8 contributions follow from the well-known expansions of the $\Gamma$ function, while
those of the sunset contribution in the limit \(\epsilon\to 0\) are encoded in the expansions
\begin{align}
J_{111}^{\frac{d}{2}}&=m_0^{2-2\eps}e^{-\eps\gamma_E}\left\{-\frac{6}{\eps^2}-\frac{9}{\eps}+\left(-\frac{21}{2}-\frac{\pi^2}{4}+c\right)+\mathcal O(\eps)\right\}\nn\\
J_{211}^{\frac{d}{2}}&=m_0^{-2\eps}e^{-\eps\gamma_E}\left\{\frac{2}{\eps^2}+\frac{1}{\eps}+\left(\frac{1}{2}+\frac{\pi^2}{12}-\frac{c}{3}\right)+\mathcal O(\eps)\right\}\nn\\
J_{222}^{\frac{d}{2}+1}&=m_0^{-2\eps}e^{-\eps\gamma_E}\left\{\frac{1}{2\eps}+\left(\frac{3}{8}-\frac{2c}{9}\right)+\mathcal O(\eps)\right\}
\end{align}
where $c=2\sqrt3\Im[\Li_2(e^{\frac{i\pi}{3}})]=3.516$, and $\gamma_E$ is the Euler-Mascheroni constant that will be absorbed in the definition of the $\overline{\text{MS}}$ renormalization scale together with the factors of $(4\pi)^\eps$ arising from the loop factors.

Eqns.~(\ref{eq:1loop}), (\ref{eq:fig8}) and (\ref{eq:sunset}) give the one and two-loop  contributions to the singularities of the effective action for the original interactions and kinetic terms, to which we shall apply the modified Minimal Subtraction ($\overline{\text{MS}}$) renormalization scheme by introducing counterterms in order to cancel such singularities.
We thus renormalize the effective action by defining  $\phi_0\equiv \sqrt{Z}\phi$, $Z\equiv 1+\delta Z$, $Zm_0^2\equiv m^2+\delta m^2$, $Z^2\lambda_0\mu^{-\epsilon}\equiv \lambda+\delta \lambda$, $Z\xi_0\equiv \xi+\delta \xi$. 
We obtain the counterterms
\begin{align}
\delta Z&=- \frac{1}{12\eps}\llf\nn\\
\frac{\delta m^2}{m^2}&=\frac{1}{\eps}\lf+\left(\frac{2}{\eps^2}-\frac{1}{2\eps}\right)\llf\nn\\
\frac{\delta \l}{\l}&=\frac{3}{\eps}\lf+\left(\frac{9}{\eps^2}-\frac{3}{\eps}\right)\llf\nn\\
\delta \xi&=\frac{1}{\eps}\left\{\frac{1}{6}+\xi\right\}\lf
+\left\{\left(\frac{1}{3\eps^2}-\frac{7}{72\eps}\right)
+\left(\frac{2}{\eps^2}-\frac{1}{2\eps}\right)\xi\right\}\llf
\label{eq:ct}
\end{align}
with $\ell\equiv \frac{\lambda}{16\pi^2}$. Observe that these counterterms are independent of $\mu$ and $m$, this provides a nontrivial cross check.
Moreover, the two-loop counterterms  coincide with the ones obtained in ref.~\cite{Toms:1982af} which also employed the background field method. However, the author stated that he was not able to apply HK methods to evaluate the sunset diagram, and instead relied on some indirect arguments using renormalizability in order to obtain the corresponding contributions.
It is precisely this lack of a systematic treatment that we tried to address in the present work.

From eq.~(\ref{eq:ct}) one obtains the well known two-loop anomalous dimension, mass anomalous dimension, and beta function of the flat theory\cite{Kleinert:2001ax},
\be
\gamma=\frac{1}{12}\llf\,,
\qquad
\gamma_m=\frac{1}{2}\lf-\frac{5}{12}\llf\,,
\qquad
\beta_\lambda=\l\left(3\lf-\frac{17}{3}\llf\right)
\ee
as well as finally the beta function for $\xi$:
\be
\beta_{\xi}=\left(\frac{1}{6}+\xi\right)\lf
-\left(\frac{7}{36}+\frac{5}{6}\xi\right)\llf
\ee

\begin{figure}
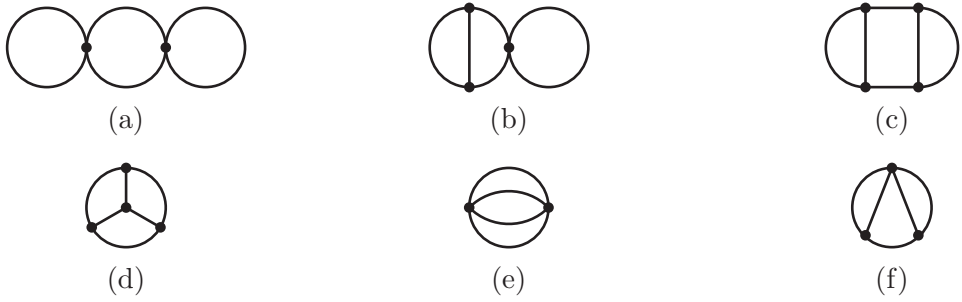

\begin{center}

\begin{subfigure}[b]{0.3\linewidth}
  \centering
  \begin{axopicture}(90,30)(0,0)
    \SetWidth{1.0}
   	\Vertex(30,15){2}
   	\Vertex(60,15){2}
	\Arc(15,15)(15,0,360)
	\Arc(45,15)(15,0,360)
	\Arc(75,15)(15,0,360)
\end{axopicture}
\caption{}
\label{fig:3bubble}
\end{subfigure}
\begin{subfigure}[b]{0.3\textwidth}
	\centering
	\begin{axopicture}(60,30)(0,0)	
	\SetWidth{1.0}
    \SetColor{Black}
	\Arc(15,15)(15,0,360)
	\Arc(45,15)(15,0,360)
	\Line(15,0)(15,30)
	\Vertex(15,0){2}
	\Vertex(15,30){2}
	\Vertex(30,15){2}
\end{axopicture}
\caption{}
\label{fig:fig8bis}
\end{subfigure}
\begin{subfigure}[b]{0.3\textwidth}
	\centering
	\begin{axopicture}(50,30)(0,0)	
	\SetWidth{1.0}
    \SetColor{Black}
	\Arc(15,15)(15,90,270)
	\Arc(35,15)(15,270,90)
	\Line(15,0)(15,30)
	\Line(35,0)(35,30)
	\Line(15,0)(35,0)
	\Line(15,30)(35,30)
	\Vertex(15,0){2}
	\Vertex(15,30){2}
	\Vertex(35,0){2}
	\Vertex(35,30){2}
\end{axopicture}
\caption{}
\label{fig:ladder}
\end{subfigure}

\begin{subfigure}[b]{0.3\textwidth}
	\centering
	\begin{axopicture}(30,40)(0,0)	
	\SetWidth{1.0}
    \SetColor{Black}
	\Arc(15,15)(15,0,360)
	\Line(15,30)(15,15)
	\Line(15,15)(2,7.5)
	\Line(15,15)(28,7.5)
	\Vertex(15,15){2}
	\Vertex(15,30){2}
	\Vertex(2,7.5){2}
	\Vertex(28,7.5){2}		
\end{axopicture}
\caption{}
\label{fig:tetra}
\end{subfigure}
\begin{subfigure}[b]{0.3\textwidth}
	\centering
	\begin{axopicture}(30,30)(0,0)	
	\SetWidth{1.0}
    \SetColor{Black}
	\Arc(15,15)(15,0,360)
	\Arc(15,0)(21.2,45,135)
	\Arc(15,30)(21.2,225,315)
	\Vertex(0,15){2}
	\Vertex(30,15){2}		
\end{axopicture}
\caption{}
\label{fig:eyeball}
\end{subfigure}
\begin{subfigure}[b]{0.3\textwidth}
	\centering
	\begin{axopicture}(30,30)(0,0)	
	\SetWidth{1.0}
    \SetColor{Black}
	\Arc(15,15)(15,0,360)
	\Line(15,30)(5,4.5)
	\Line(15,30)(25,4.5)
	\Vertex(15,30){2}
	\Vertex(5,4.5){2}
	\Vertex(25,4.5){2}		
\end{axopicture}
\caption{}
\label{fig:last}
\end{subfigure}
\caption{Three-loop diagrams.}
\label{fig:3loop}
\end{center}
\end{figure}

We stress that the extension of the calculation to three loops is in principle straightforward. Here we content ourselves to a brief outline. The three-loop diagrams are given in figure \ref{fig:3loop}. All but the diagrams in figure \ref{fig:3bubble} and \ref{fig:eyeball}  require the knowledge of the full off-diagonal HK $B(\t,y,y')$. Take, for instance, the diagram in figure \ref{fig:fig8bis}. 
The ingredients for the evaluation of eq.~(\ref{eq:master3}) read
\begin{align}
\Delta(\t_i)={}&\t_5\left[\t_1\t_2+(\t_3+\t_4)\t_1+(\t_3+\t_4)\t_2\right]\nn\\
I'(\t_i,p_{n'})={}&\frac{1}{(4\pi)^\frac{3d}{2}\Delta^\frac{d}{2}}e^{\frac{\t_5}{\Delta}\left[\t_4(\t_1\t_2+\t_2\t_3+\t_1\t_3)p_2^2
+\t_3(\t_1\t_2+\t_2\t_4+\t_1\t_4)p_3^2-2\t_3\t_4(\t_1+\t_2)p_2p_3 \right]}\nn\\
\Gamma(\t_i,y_n)={}&-\lambda_0^3\sqrt{g(y_2)g(y_3)}\phi_0(y_2)\phi_0(y_3)\times\nn\\
&\times B(\t_1,y_2,y_3)B(\t_2,y_2,y_3)B(\t_3,y_2,0)B(\t_4,y_3,0)B(\t_5,0,0)
\end{align}
and the diagram has a symmetry factor $\frac{1}{8}$. The computation of $\mathcal L_{\rm eff}$ from this diagram is just as straightforward as for the two-loop case. Concerning the integrals over Schwinger parameters, diagrams of figure \ref{fig:3bubble} and  \ref{fig:fig8bis} reduce to corresponding integrals of lower loop order, while the diagrams in figures \ref{fig:ladder} to \ref{fig:last} involve genuine three-loop integrals (see, for instance, ref.~\cite{Martin:2016bgz}).

\section{Conclusions}
\label{sec:conclusions}

We have developed a formalism for the calculation of multi-loop Feynman graphs of scalar fields in curved spacetime. 
The formalism relies on the expansion of all quantities, in particular the heat kernel in Riemann normal coordinates. 
The result is given in closed from in terms of graph polynomials (depending on the Feynman-Schwinger parameters).
 This leaves as the only remaining step the calculation of the integrals over Schwinger parameters, which are equivalent to the evaluation of scalar $L$ loop vacuum graphs. The latter are fully known up to three loop order \cite{Martin:2016bgz}, and powerful methods exist for the general case (see for instance ref.~\cite{Weinzierl:2022eaz}). 

Our formalism could be easily generalized in various ways. Firstly, one may wish to include fields of nonzero spin. 
Fermionic heat kernels will depend on the propagator momenta \cite{vonGersdorff:2022kwj,vonGersdorff:2023lle}, which means that one can straightforwardly include them by going from the special case eq.~(\ref{eq:master2}) back to the general one eq.~(\ref{eq:master}).
Secondly one may wish to include arbitrary non-Abelian background gauge fields, which within our formalism is most naturally done by employing the radial gauge along the lines of ref.\cite{Luscher:1982wf}. We leave these directions to future work.

\section*{Ackknowledgements}
GG acknowledges financial support by the Conselho Nacional de Desenvolvimento Científico e Tecnológico (CNPq) under fellowship number 309448/2020-4.

\appendix

\section{Conventions and notation}
\label{sec:conventions}

Our metric has signature $+---$. Our sign conventions amount to  
\begin{align}
\Gamma^\mu_{\r\s}&=\tfrac{1}{2}g^{\m\a}(g_{\r\a,\s}+g_{\s\a,\r}-g_{\r\s,\a})\,\nn\\\
R^\m_{\ \n\r\s}&=\Gamma^\m_{\n\s,\r}-\Gamma^\m_{\n\r,\s}+\dots
\end{align}
For the Ricci tensor and scalar curvature, we use $R^{}_{\mu\nu}=R^\a_{\ \m\a\n}$, $R=R^{\m}_{\ \m}$.
Semicolons before indices denote covariant differentiation, while commas denote partial differentiation. For biscalar functions (and similarly for bitensors) we use the notation
\be
f_{;\mu\nu\dots;\rho\sigma\dots}(y,y')
	\equiv
	(\dots\nabla_\n\nabla_\m)(\dots\nabla'_\s\nabla'_\r)
	f(y,y')
\ee
If no derivatives wrt the first argument appear, this results in a double semicolon, 
e.g.~$f_{;;\r\s\dots}(y,y')$.
A similar notation holds for partial derivatives, in which case we extend this to general functions of two arguments (not necessarily transforming as biscalars).

\section{Riemann Normal Coordinates}
\label{sec:rnc}

For convenience, we reproduce here some relevant expansions of geometric quantities in RNC's (see for instance ref.~\cite{willmore1993riemannian,Brewin:2009se})

\begin{itemize}
\item
Metric, inverse metric, metric determinant, connection.
\begin{align}
g_{\mu\nu}(y)&=\eta_{\mu\nu}-\tfrac{1}{3}R_{\m\a\n\b}y^\a y^\b
	-\tfrac{1}{6}R_{\m\a\n\b;\g}y^\a y^\b y^\g 
	\nn\\
	&+\left(\tfrac{16}{15}\eta^{\r\s}R_{\m\a\r\b}R_{\n\g\s\d}
		-\tfrac{6}{5}R_{\m\a\n\b;\g\d}    
		 \right)\tfrac{1}{24}y^\a y^\b y^\g y^\d \\
g^{\mu\nu}(y)&=\eta^{\mu\nu}
	+\tfrac{1}{3}R^{\m \ \n}_{\ \a\ \b}y^\a y^\b
	+\tfrac{1}{6}R^{\m \ \n}_{\ \a\ \b;\g}y^\a y^\b y^\g 
	\nn\\
	&+\left(\tfrac{8}{5}\eta^{\r\s}R^{\m}_{\ \a\r\b}R^{\n}_{\ \g\s\d}
		+\tfrac{6}{5}R^{\m\ \n}_{\ \a\ \b;\g\d}    
		 \right)\tfrac{1}{24}y^\a y^\b y^\g y^\d \\
[ g(y)]^q&=1-\tfrac{q}{3}R_{\m\nu}\,y^\m y^\n
	-\tfrac{q}{6}R_{\m\n;\r}y^\m y^\n y^\r
	\nn\\&	
	+\left(\tfrac{q^2}{18}R_{\m\n}R_{\r\s}
	-\tfrac{q}{20}R_{\m\n;\r\s}-\tfrac{q}{90}R^{}_{\mu \a \nu \b}R_{\r\ \s}^{\ \a\ \b}
		\right)y^\m y^\n y^\r y^\s+\dots
	\\
\Gamma^\l_{\m\n}(y)&=-\tfrac23R^{\lambda}_{\ (\mu\nu)\rho}\, y^\r
+	\left(\tfrac{1}{12}R^\lambda_{\ \r\s(\mu;\nu)}
	-\tfrac{5}{12}R^\lambda_{\ (\mu\nu)\r;\s}\right)y^\r y^\s+\dots
\label{eq:ChristoffelRNC}
\end{align}
\item
Scalar functions and
general covariant tensors 
\begin{align}
\phi(y)&=\sum_{n=0}^\infty \frac{1}{n!}\phi_{;\mu_1\dots\mu_n}y^{\m_1}\cdots y^{\m_n}
\nn\\
T_{\l_1\dots \l_r}(y)&=T_{\l_1\dots \l_r}+T_{\l_1\dots \l_r;\m}\,y^\m
\nn\\
&+\tfrac{1}{2}\left(T_{\l_1\dots \l_r;\m\n}+\tfrac{1}{3}\sum_{a=1}^r R^\a_{\ \m\n\l_a}T^{}_{\l_1\dots \l_{a-1}\a \l_{a+1}\dots \l_r}\right)y^\m y^\n+...
\end{align}
\item
Synge's world function
\begin{align}
\sigma(y,y')=&
\ \tfrac{1}{2}g_{\mu\nu}\,(y-y')^\mu(y-y')^\nu -\tfrac{1}{6}R_{\mu\r\nu\s}\,y^\mu y^\nu y'^\r y'^\s 
\nn\\
	&-\tfrac{1}{12}R_{\mu\r\nu\s;\t}\, 
		y^\mu y^\nu (y^\t+y'^\t) y'^\r y'^\s
\nn\\
&
	-\left(
	\tfrac{8}{15}R^{\alpha}_{\ \k\m\r}R_{\alpha\t\n\s}^{}
	+\tfrac{2}{5}R_{\mu\r\nu\s;\k\t}\right) 
		\tfrac{1}{48}\, y^\m y^\n (y^\k y^\t+y'^\k y'^\t) y'^\r y'^\s
	\nn\\&	 
	-\left(
	\tfrac{8}{5}R^{\alpha}_{\ \mu\nu\r}R_{\alpha\s\t\k}^{}
	+\tfrac{3}{20}R_{\mu\r\nu\s;\k\t}
	+\tfrac{3}{20}R_{\mu\r\nu\s;\t\k}
	\right)
		\tfrac{1}{36}\,y^\mu y^\nu  y^\k y'^\r y'^\s y'^\t 
+\dots
\label{eq:SyngeRNC}
\end{align}
\end{itemize}

\section{Symmetric treatment of the vertices.}
\label{sec:jacobian}

In this appendix we give an alternative formalism, which treats the $V$ vertices in a completely symmetric way. This is achieved by choosing as the reference point (around which the background fields are expanded) the center of mass point of the vertices.
Given $V$ vertex points $x_n$, we define $\bar x$ to be the extremum  \footnote{In Euclidean signature this is a local minimum.} of the function $\chi(x)\equiv\sum_{n=1}^V \sigma(x,x_n)$ 
where $\sigma$ is Synge's world function, equal to one half the geodesic distance squared between its two arguments.\footnote{A maybe simpler but less symmetric choice $\bar x=x_1$ (or any other vertex) is discussed at the end of this section.}
Cleary, for any given set of points $x_n$, $\chi(x)$ is a scalar function of $x$ and hence its extremum transforms covariantly under coordinate changes. 
In RNCs with base point $x$, $\delta(x,\bar x)$ simply becomes $\delta(\bar y)$. Moreover, in the flat limit, $\bar y$ reduces to $\bar y_0$ of eq.~(\ref{eq:ybar}).
We will show below that $\bar y=0$ if and only if $\bar y_0=0$, and hence we can write
\be
\delta(\bar y)=\delta(\bar y_0)D(y_n)
\ee
where $D(y_n)$ is a Jacobian whose RNC expansion we derive explicitly below. Therefore, we can use the original function $I(\t_i,p_n)$ in our master formula eq.~(\ref{eq:master3}), provided we include in the definition of $\Gamma$ the Jacobian, $\Gamma(\t_i,y_n)\to 
D(y_n)\Gamma(\t_i,y_n)$. 

In the remainder of this appendix, we calculate the Jacobian $D(y_n)$.
The RNCs of the center of mass point are  calculated from 
\be
\sum_n\sigma_{,\mu}(\bar y,y_n)=0
\label{eq:ybardef}
\ee
When the first argument of $\sigma$ coincides with the base point of the RNCs, we have
$\sigma(0,y)=\frac{1}{2}y^2$ and hence
\be
\sigma_{,,\mu}(0,y)=\eta_{\mu\nu}y^\nu\qquad
\sigma_{,,\mu\nu}(0,y)=\eta_{\mu\nu}
\label{eq:sigmamu2}
\ee
Slightly less trivial, we also have (see e.g.~eq.~(3.3) of reference \cite{Poisson:2011nh})
\be
\sigma_{,\mu}(0,y)=-\eta_{\mu\nu}y^\nu
\qquad
\sigma_{,\mu,\nu}(0,y)=-\eta_{\mu\nu}
\label{eq:sigmamu1}
\ee
This relation is crucial, as it implies that  $\bar y=0$ is a solution to  eq.~(\ref{eq:ybardef}) if and only if $\bar y_0=0$.
Therefore, we can write
\be
\delta(\bar y)=\left|\det \frac{\partial \bar y^\mu}{\partial \bar y_0^\nu}\right|^{-1}\delta(\bar y_0)
\ee
One has to be a bit careful here because the $y_n$ and $\bar y_0$ are not independent.
One way to deal with it is to temporarily make a change of variables from, say, $y_V$ to $\bar y_0$. We can then compute the Jacobian directly from differentiating eq.~(\ref{eq:ybardef}).
One obtains
\be
0=
\sum_{n=1}^V \sigma_{,\mu\sigma}(\bar y,y_n)\frac{\partial\bar y^\sigma}{\partial \bar 
y_0^\rho}+\sigma_{,\mu,\sigma}(\bar y,y_V)\frac{\partial y_V^\sigma}{\partial \bar 
y_0^\rho}
\ee
Putting $\bar y=\bar y_0=0$ gives
\be
\sum_n \sigma_{,\mu\sigma}(0,y_n)
\left.\frac{\partial\bar y^\sigma}{\partial \bar y_0^\rho}\right|_{y_0=0}
-V\eta_{\mu\rho}=0
\ee
where we used Eq.~(\ref{eq:sigmamu1}).
Hence, we can write
\be
\delta(\bar y)=\left|
\det\left(\frac{1}{V}\textstyle\sum_n \sigma_{,\mu\sigma}(0,y_n)\right)
\right|\delta(\bar y_0)
\equiv D(y_n)\delta(\bar y_0)
\ee
The expansion in RNCs of $\sigma_{,\mu\nu}(0,y)$ can be found from eq.~(\ref{eq:SyngeRNC}), yielding for the Jacobian \footnote{
We notice without proof that it is possible to relate the expansion coefficients of $\sigma_{,\d\l}(0,y)$ at arbitrary order $m$ to the expansion coefficients of the inverse metric:
\be
\sigma^{}_{,\d\l,\mu_1\dots \mu_m}(0,0)=
-\frac{1}{m-1}
\eta_{\d\mu}\eta_{\l\nu}g^{\mu\nu}_{\ \ ,\mu_1\dots \mu_m}(0)
\ee
}
\bea
D(y_n)
&=&1-\tfrac{1}{3}R_{\mu\nu}\, \bar y_0^{\mu\nu}
-\tfrac{1}{12}R_{\mu\nu;\rho}\, \bar y_0^{\mu\nu\rho}
+\left(-\tfrac{1}{60}R_{\mu\nu;\rho\sigma}
-\tfrac{1}{45}
R_{\a\mu \g\nu}R_{\b\rho \d\sigma}\eta^{\a\b}\eta^{\g\d}
\right)\bar y_0^{\mu\nu\rho\sigma}\nn\\
&&
+\ \tfrac{1}{18}\left(
R_{\mu\nu}R_{\rho\sigma}-
R_{\a\mu \g\nu}R_{\b\rho \d\sigma}\eta^{\a\b}\eta^{\g\d}
\right)\bar y_0^{\mu\nu}\bar y_0^{\rho\sigma}+\ \mathcal O(R^\frac{5}{2})
\eea
where we defined the totally symmetric tensors
\be
\bar y_0^{\, \nu_1\dots \nu_m}\equiv \frac{\sum_n y_n^{\nu_1}\cdots y_n^{\nu_m}}{V}
\ee

\bibliographystyle{unsrt}

\bibliography{literature}

\end{document}